\documentclass[a4paper,fleqn]{cas-sc}

\usepackage[authoryear]{natbib}

\def\tsc#1{\csdef{#1}{\textsc{\lowercase{#1}}\xspace}}
\tsc{WGM}
\tsc{QE}
\tsc{EP}
\tsc{PMS}
\tsc{BEC}
\tsc{DE}

\begin{document}
\let\WriteBookmarks\relax
\def\floatpagepagefraction{1}
\def\textpagefraction{.001}

\shorttitle{Surface Charging in Meteorite Craters on Asteroids}

\shortauthors{Zhiying Song et~al.}

\title [mode = title]{Three-dimensional Simulation of Surface Charging in Meteorite Craters on Rotating Asteroids}                      
\tnotemark[1]

\tnotetext[1]{This work was supported by the National Natural Science Foundation of China (grant No.42241148 and No.51877111).}

%
\author[1]{Zhiying Song}[orcid=0000-0003-3498-2172]


\ead{songzy815rp@163.com}


\credit{Conceptualization of this study, Methodology, Data curation, Software, Writing - Final draft compilation}

\author[1]{Zhigui Liu}[orcid=0000-0002-9859-0277]

\ead{liuzhiguiyyy@163.com}

\credit{Methodology, Data curation, Literature review, Writing - Original draft preparation}

\author[1]{Ronghui Quan}[orcid=0000-0002-4321-9008]

\ead{quanrh@nuaa.edu.cn}

\cormark[1]

\credit{Software, Writing - Review and editing}

\cortext[cor1]{Corresponding author}

\affiliation[1]{organization={College of Astronautics, Nanjing University of Aeronautics and Astronautics},
    city={Nanjing},
    postcode={210016}, 
    state={Jiangsu},
    country={China}}

\begin{abstract}
Meteorite craters on the asteroid surface obstruct the horizontal flow of solar wind, forming a plasma wake that modulates the particle fluxes and the electrostatic environment far downstream. In this study, surface charging properties of asteroids with nontrivial terrain are simulated based on neural network and the finite element method. Key factors such as the location, size and depth-to-width ratio of craters are all considered. Under normal conditions, as the latitude of the crater increases, the potential variation at its floor during a rotation gradually becomes smoother, finally stabilizing around -3V with minor fluctuations as the crater approaches the poles. For craters with a depth-to-width ratio greater than 0.5, because of the diverging motions of electrons and the less deflected trajectories of ions, completely different charging results are observed under parallel and perpendicular solar wind incidence, the potential around the crater floor decreases and increases with the rising depth-to-width ratio, respectively. While the surface potential appears indifferent to changes in crater size, only during solar storms, the floor of large-scale craters, such as those with a diameter of 800m, perform a 9.13V decrease in potential compared to small craters of 50m. Both studies of localized plasma flow field and the surface charging phenomenon of asteroids have substantial influence on the future safe landing and exploration missions.


\end{abstract}


\begin{highlights}
\item Proposed a method based on neural networks and finite element analysis, which enables efficient and accurate 3D dynamic simulation of asteroids.
\item Comparative analysis of surface charging results for craters with different characteristics, including location, size, and depth-to-width ratio.
\item Studying the surface charging properties of craters during the solar storm and at various solar wind incidence angles.
\end{highlights}


\maketitle

\section{Introduction}

The asteroid surface that is in contact with the solar wind plasma collects electric charges of impinging electrons and ions, resulting in photoelectric emission and secondary electron emission, then generate currents and cause the asteroid surface to exhibit an observable potential. This surface charging phenomenon, primarily driven by the photoelectric effect, electron together with ion attachment and secondary electron emission, will interact with the surrounding plasma, changing the density distribution and temperature of the electrons and ions. Therefore, the asteroid's surface potential is the result of the coupling between the electric field and the plasma. However, within a shadowed crater, the direct flow of the solar wind is obstructed by the upstream crater wall, forming a complex plasma wake with spatially inhomogeneous plasma and strong secondary currents. The surface potential in this area will show significant differences compared to other flat regions, even further affecting the surface charging results of the entire asteroid.

The theoretical foundation of surface charging is relatively well-established, and related research can be divided into three main areas: the electrification properties and electrostatic driving dynamics of surface dust particles, the overall charging properties of asteroid surfaces, and deep charging of asteroids induced by surface charging. Studies in the first category center on the charged properties \citep{2016JGRE..121.2150Z,2019ITPS...47.3710O} and the floating motion \citep{hartzell2013dynamics} of regolith grains on the surface, mainly use models such as magnetohydrodynamic (MHD) or hybrid particle-in-cell (Hybrid PIC). We focus on the second category. On the dayside of the asteroid, the photoelectron current usually dominates, resulting in a consistently positive surface potential, and form a photoelectron plasma sheath above the surface, theoretical possibilities have been discussed \citep{2020Ap&SS.365...23K}. Correspondingly, on the nightside, the charging current mainly originates from the incidence of electrons and ions in the solar wind plasma, due to the higher differential influx of electrons compared to ions, the surface presents a negative potential, which even reach thousands of volts \citep{2007GeoRL..34.2111H} during solar energetic particle (SEP) events. However, if sufficient secondary electrons are emitted, the nightside can be charged positively \citep{halekas2009lunar}. The maximum electric field strength, approximately 10 times that on the dayside \citep{2023ApJ...952...61X}, appears at the terminator where there is a transition from sunlight-driven positive to plasma-created negative potential \citep{farrell2007complex}, with electron temperature being the most critical parameter \citep{2014P&SS...90...10S}. Considering the densely distributed and irregular impact craters on the asteroid surface, localized areas can exhibit unconventional charging scenarios. At the terminator, intense electric fields are observed around small-scale topographic features such as ridges and depressions. Electrons with high thermal velocities readily penetrate into the shadowed regions of these craters, while protons require much higher velocities to reach such areas. This results in the formation of a pronounced local negative potential, which is largely determined by the slope of the shadowed surfaces within the depressions \citep{borisov2006charging}.

Although the charging characteristics of complex topography near the terminator have been analyzed, static simulations of irregularly shaped asteroids at specific angles are also feasible \citep{2014Icar..238...77Z}. However, how crater's characteristics, such as location, size, and depth-to-width ratio, influence surface charging of rotating asteroids remains an urgent problem to be solved. In this study, neural networks and the finite element method (FEM) were used to dynamically simulate the surface potential of asteroids with meteorite craters, greatly improving the efficiency of real-time analysis. We used COMSOL Multiphysics to model the asteroid and perform transient simulations of its charging process. We first explored asteroids exposed to typical solar wind, considering craters distributed across different latitudes, from the equator to the poles. Subsequently, when we study the impact of crater size and depth-to-width ratio on surface charging, the solar storm is taken into consideration. Analyzing the surface charging characteristics of asteroids with complex terrain under the wide variety of conditions, especially rotating asteroids, will contribute to understanding the formation and evolution of asteroid surface morphology, and feed forward into future in asteroid landings and in-situ exploration missions.

Details of the establishment of the model and initial value configuration are shown in Section \ref{sec:MD}, including the description of our method. The parameter settings and analysis of simulation results are present in Section \ref{sec:Sim}. Finally, we summarize our study in Section \ref{sec:Con}.

\section{Model Description} \label{sec:MD}

A three-dimensional model was developed using COMSOL Multiphysics to investigate the surface charging phenomenon in solar wind plasma. Our simulation integrates neural networks and the finite element method (FEM). The neural network outputs the asteroid's surface potential based on the surrounding plasma, while FEM is responsible for the electric field and plasma environment in the outer space around the asteroid. These two parts feed into each other at each time step.

BP neural network is a multi-layer feedforward network trained with error inverse propagation algorithm \citep{zhu2023method}. Its topology can be divided into three levels: input layer, hidden layer, and output layer \citep{adil2022effect}. Take mean square error (MSE) as standard, neural network learns the input-output dataset through the steepest gradient descent method, via reverse propagation to continuously adjust the weights of each layer to minimize errors. It can structure and store massive mapping relationships hidden within a large dataset, and perform excellent generalization ability and strong fault tolerance \citep{2022MRE.....9b5504L}.

We trained two neural networks, in order to calculate the instantaneous potential and equilibrium potential, respectively. The latter is mainly used for static qualitative analysis of asteroid charging properties. 

The charging currents on the asteroid surface include photoelectron currents, incident electron and ion currents, backscattered currents, secondary electron currents, and conduction currents. The photoelectric effect, electron and ion attachment, and secondary electron emission are the primary mechanisms driving surface charging. Moreover, in addition to solar wind conditions, the mineral composition of the asteroid surface also counts. For instance, the photoelectron current is dependent on the work function, while conductivity impacts conduction currents. Therefore, it is important to consider these aspects when selecting input parameters of the neural network. In this study, the input parameters for the neural network can be divided into three categories: solar wind parameters, including electron and ion density, electron and ion temperatures, and solar radiation; material parameters, including work function, maximum secondary electron emission yield and the corresponding energy, relative permittivity, and conductivity; current potential and the simulation time step. Notably, the third category is only adopted for calculating instantaneous potential. The training dataset is generated from numerical simulations\citep{2004sei..book.....H,Impact,7407629}. For all the testing data, the errors between potentials output by neural network and those calculated by numerical simulation are within 5$\%$. This demonstrates that the neural network can effectively compute the surface potential of asteroids under various conditions.

We used FEM to solve the plasma environment. Its core idea is to discretize a complex continuum or system into a set of smaller, simpler elements for numerical computation. We considered the entire simulation region to be composed of many interconnected small elements called finite elements. Each element is a tetrahedron, all the elements connect through nodes and fit together without gaps, with the edges being common between neighboring elements \citep{2024JPhCS2701a2105C}. Thus, solving the partial differential equations (e.g. the drift-diffusion equations for electrons and ions) for the entire region can be divided into solving of these small regions, thereby improving the accuracy of local computations.

Our simulation can be summarized as the following steps: a. Input numerous solar wind and material parameters, compute instantaneous and equilibrium potentials on the asteroid surface under these conditions with numerical integration, thereby building a dataset; b. Divide the dataset into training and testing sets, train and optimize the neural network; c. Model the asteroid and initialize the electrostatic field of the asteroid surface and surrounding plasma; d. Input parameters (including solar wind plasma, solar irradiation, and surface material) at various locations into the neural network, and output the surface potential; e. Calculate the potential and electric field distribution within the remaining domain; f. Simulate the drift-diffusion motion of electrons and ions using FEM, then compute their density and temperature; g. Update the input parameters of the neural network, and output the surface potential at the next simulation time step; h. Iterate steps d-g until simulation time reaches the required duration.

\begin{figure}[ht!]
\centering
\includegraphics[width=0.8\linewidth]{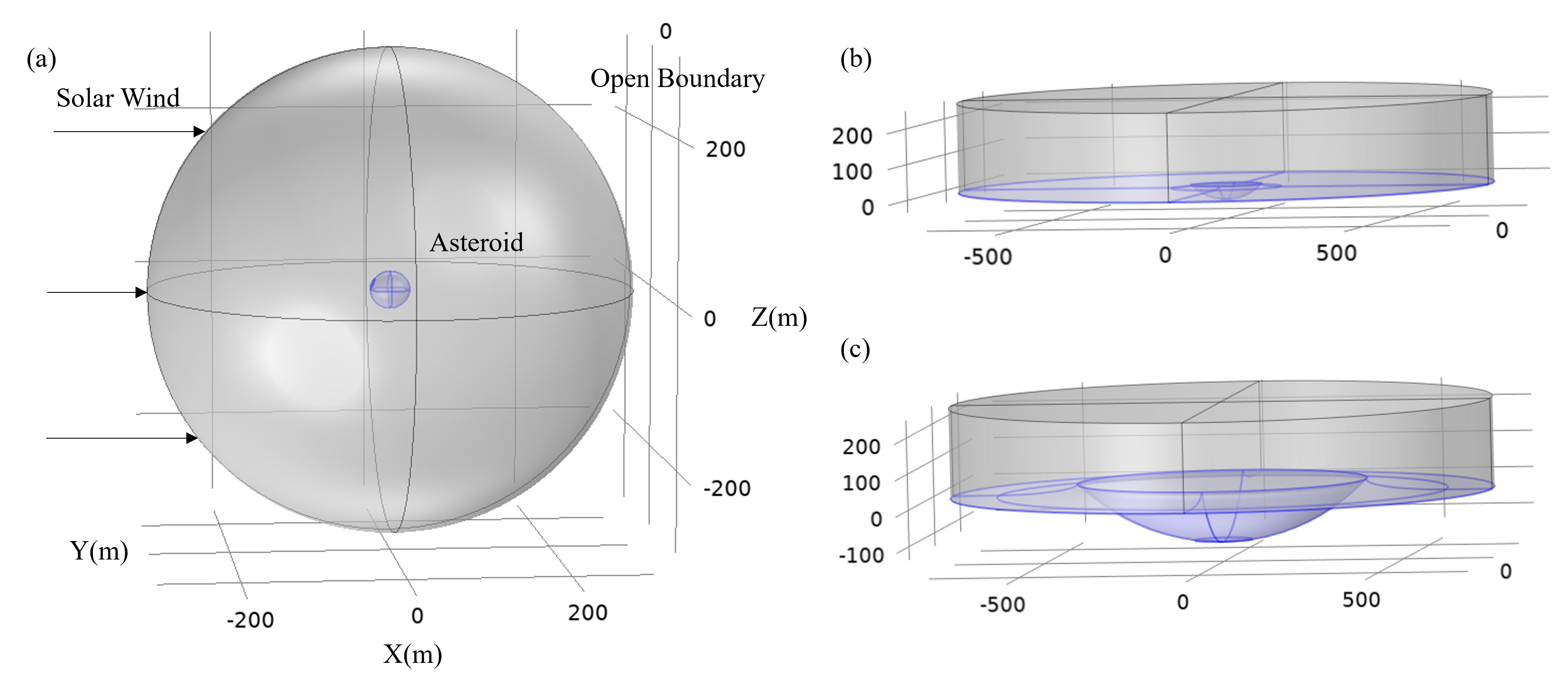}
\caption{The schematic diagram of the simulation geometries. (a): the gray sphere indicates our simulation domain and the blue geometry represents the asteroid with meteorite crater. The three black arrows on the left show the direction of solar wind. (b) and (c): the blue areas represent the asteroid surface.
\label{fig:Fig1}}
\end{figure}

In the first part, which studies the effect of crater's location on charging results, the entire asteroid was modeled. We used the shape of Ceres to represent the asteroid \citep{2023RemS...15.4209N}. As indicated in Figure \ref{fig:Fig1} (a), the simulation domain is a sphere with a radius of 300m, the asteroid locates at its center, with the crater positioned at a latitude of 30$^{\circ}$N. The Z-axis is parallel to the spin axis of the asteroid, and the solar wind flows in the +X direction, electrons and ions are emitted from the outer boundaries. For the part which focuses on the size of the meteorite crater, we modeled the area 750m around the crater. Figures \ref{fig:Fig1} (b) and (c) illustrate the modeling of craters with diameters of 200m and 800m, respectively. A meteorite crater typically consists of several features, including the crater rim, wall, and floor. The rim is the raised area surrounding the crater, formed by the accumulation of material ejected during the impact event, while the wall is usually steep, created as large amounts of debris are displaced, and the floor tends to be more flat. The morphology of the craters is defined by algebraic equations \citep{lund2020fully}. We use cylindrical shapes to represent craters when we study the influence of depth-to-width ratio.

\section{Simulation} \label{sec:Sim}

We conducted dynamic three-dimensional simulations of surface charging under different conditions, categorized into the following three main categories.

\subsection{Location}
\label{subsec:Loc}

The number density of solar wind is $\mathit{\rm 5\times10^6/m^3}$, with electron and ion temperatures of 8.6eV. Electrons and ions are emitted from the outer boundaries with a bulk speed of 450km/s, an open boundary condition is applied here, allowing all particles to escape through it. Secondary electron emission is also considered.  Plagioclase is regarded as the material on the surface of asteroid, with a work function of 5.58eV, and a threshold wavelength of 238nm (sourced from a project summary report of the National Natural Science Foundation of China, No.41572037).

Surface charging of seven asteroids, positioned from the equator to the pole, were simulated, with each asteroid having a rotation period of one day. In the initial state of our simulation, the meteorite crater is positioned on the dayside of the asteroid, and the solar wind is incident perpendicularly into the crater.

\begin{figure}[ht!]
\centering
\includegraphics[width=0.9\linewidth]{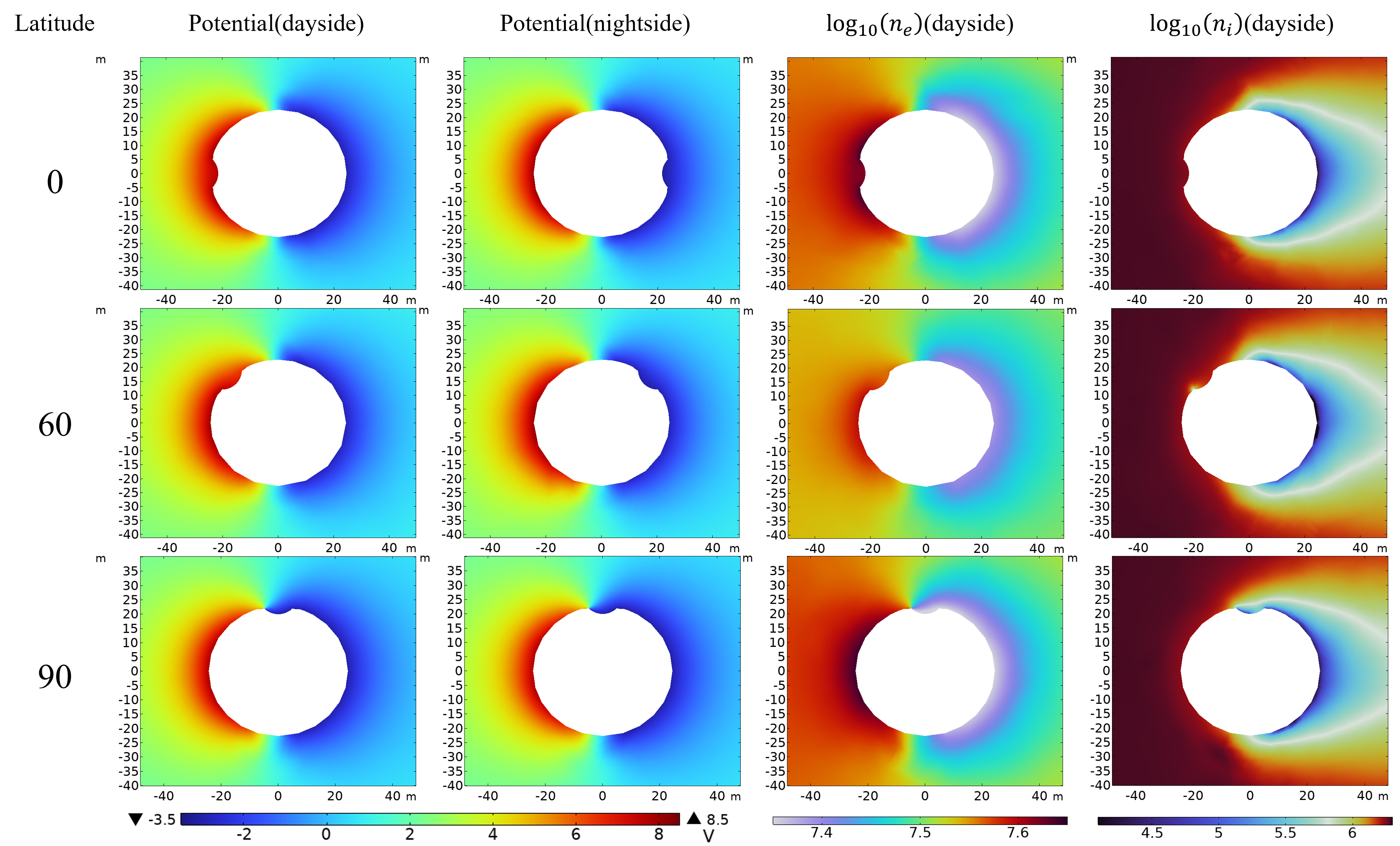}
\caption{Surface potential and surrounding plasma of the asteroid in the X-O-Z plane. Subfigures in the first column show surface potentials of asteroids with the crater on the dayside, those in the second column display potentials when asteroids have rotated half of its period, i.e. the crater is on the nightside. Other two columns represent densities of electrons and ions around asteroid respectively.
\label{fig:Fig2}}
\end{figure}

Figure \ref{fig:Fig2} displays the surface charging results of three asteroids with meteorite craters, as well as the distribution of electron and ion densities around them. According to the kinetic theory of plasma \citep{1978npi..book.....G,1970pewp.book.....G}, electrons respond more promptly to the electric field because of their much higher mobility and diffusion coefficient. Therefore, they have a relatively uniform distribution, there will not be exaggerated faults in density behind asteroid. Correspondingly, ions, primarily driven by the general flow of the solar wind, form a visible plasma wake at the tail of the asteroid.

\begin{figure}[ht!]
\centering
\includegraphics[width=0.9\linewidth]{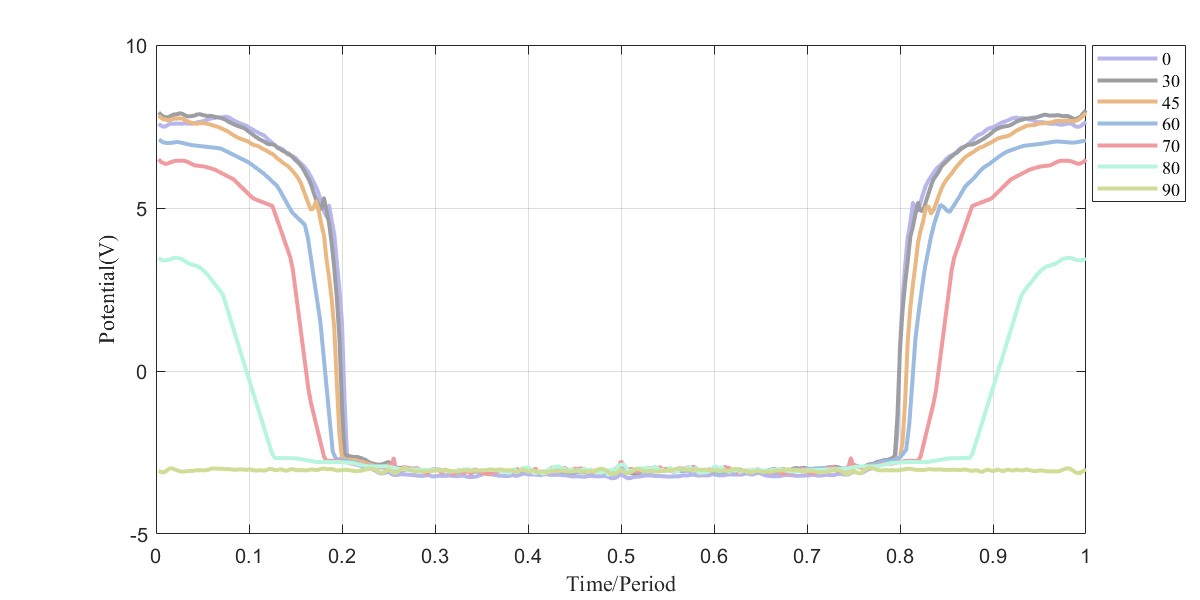}
\caption{Potential at the floor of craters located at different latitudes. Since the results of the southern hemisphere are approximately equal to those of the northern hemisphere, only the northern hemisphere results are presented.
\label{fig:Fig3}}
\end{figure}

Regarding the surface charging around meteorite craters, Figure \ref{fig:Fig3} shows the results at the floor of them. When the crater rotates to the nightside, because of the shielding effect of the asteroid against the solar wind, the potentials of craters at different latitudes show no significant variation, remaining around -3V with only slight fluctuations due to the rotation of asteroids. On the dayside, however, the potentials exhibit more pronounced changes due to direct solar wind exposure. The solar radiation received by the surface depends on the solar elevation angle, it peaks at the subsolar point, and gradually decreases as the solar elevation angle decreases, eventually reaching zero on the nightside after crossing the terminator. Additionally,  at higher latitudes, because of the obstruction caused by upstream topography, the bottom of the crater stays in shadow, resulting in lower maximum potential. This effect becomes gradually pronounced with increasing latitude. Given that the modeled crater in this section has a relatively small depth-to-width ratio of 0.21, its floor remains exposed to sunlight directly at latitudes below 45$^{\circ}$. However, when the crater's latitude exceeds 60$^{\circ}$, the maximum potential at its floor decreases sharply with increasing latitude, dropping from 7.10V at 60$^{\circ}$ to 3.47V at 80$^{\circ}$, and eventually stabilizing around -3V when the crater reaches the pole. This phenomenon is caused by the combined effects of reduced solar radiation and increased shadowing from the crater walls.

\subsection{Size}
\label{subsec:Siz}

Crater sizes can vary widely, from just a few meters to several hundred kilometers. In this study, the rotation period of the asteroid is one hour, four craters with diameters of 50m, 200m, 500m and 800m were modeled, each with a depth-to-width ratio of 0.21.

\begin{figure}[ht!]
\centering
\includegraphics[width=0.8\linewidth]{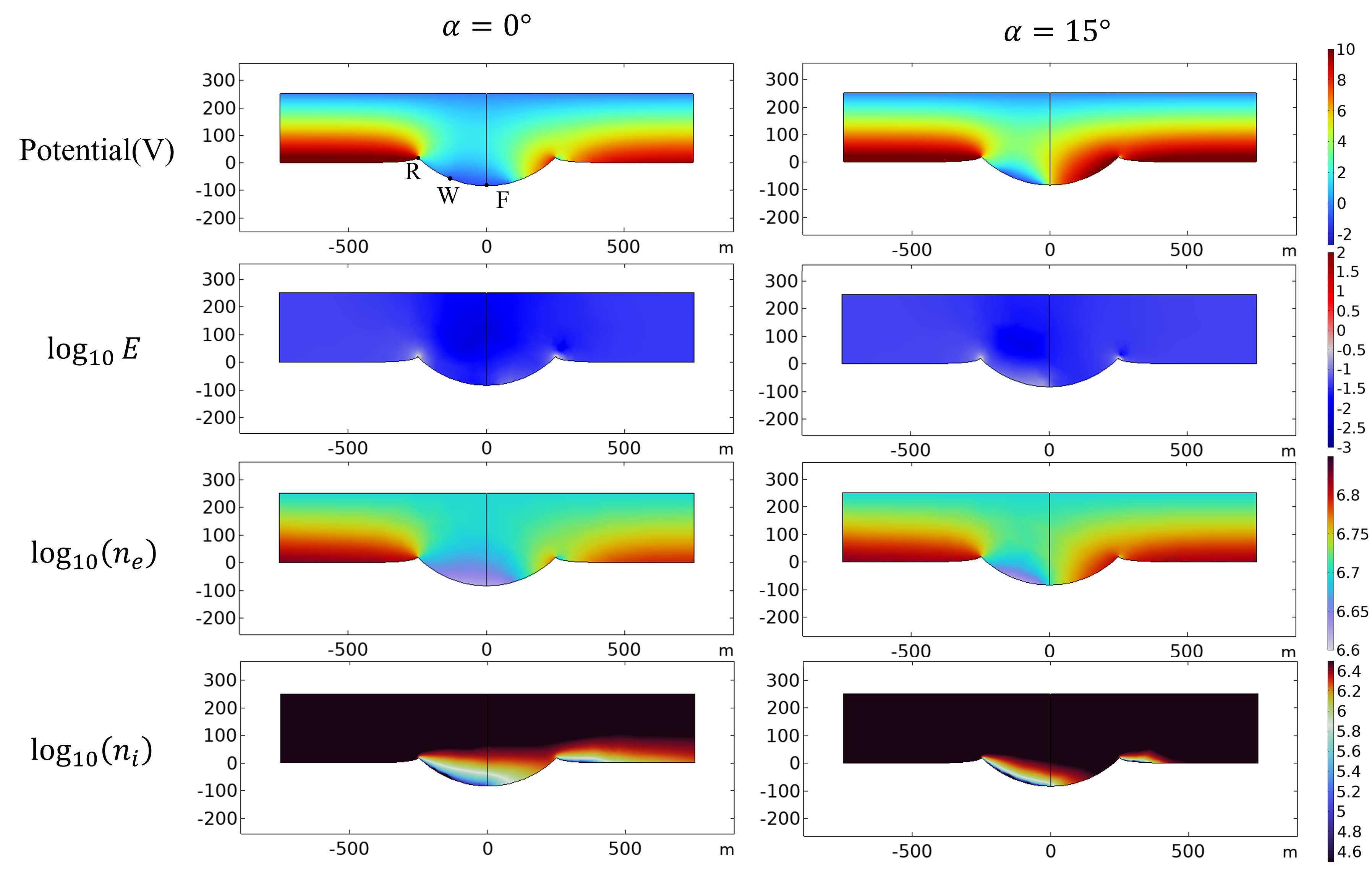}
\caption{Charging results in the X-O-Y plane for the crater with a diameter of 500m. $\alpha$ represents the angle between the plane of the crater and the incident direction of the solar wind, where $\alpha=0^{\circ}$ indicates that the solar wind is horizontally incident from the +X direction.
\label{fig:Fig4}}
\end{figure}

The results for the crater with diameter of 500m under typical solar wind conditions, with parameters provided in Section \ref{subsec:Loc}, are presented in Figure \ref{fig:Fig4}. Near the crater rim, the rapid alternation between illuminated and shadowed areas over a short distance, along with the swift transition from maximum positive potential to negative potential, strengthens the surrounding electric field, which can reach 63.07V/m at point $R$ when the solar wind is incident horizontally, specifically, the crater is near the terminator.

Electrons race into the crater ahead of the slower ions, forming a
persistent non-neutral cloud of negative space charge. For instance, at the point $W$ in the plasma wake, the electron density can reach $\mathit{\rm 4.23\times10^6/m^3}$, while the ion density is only $\mathit{\rm 5.84\times10^4/m^3}$. The low ion density decreases the frequency of electron collisions and scattering, leading to an increase in electron temperature. Additionally, the absence of photoelectron current causes electron and secondary electron currents to become the dominant contributors to surface charging, resulting in shadowed areas exhibiting the minimum potential.

According to the canonical theory of quasineutral plasma expansion \citep{1979PhFl...22.2300M}, the ion density in the crater is expressed as the product of the planar density and an exponential function of a "self-similar" dimensionless ratio \citep{2012JGRE..117.0K03Z}, which denoted as $\xi= \frac{y_c}{x_c} \cdot \frac{v_{s w}}{c_{s}}$, where $y_c$ and $x_c$ are the vertical and horizontal distance relative to point $R$, respectively, $v_{s w}$ is the solar wind convection speed, and $c_{s}$ is the ion sound speed. Therefore, the density difference caused by crater wall shielding is only related to the ratio of the vertical and horizontal distances from the current position to $R$. Consequently, for craters with the same depth-to-width ratio but different sizes, the charging results generally exhibit minimal variation under normal conditions.

\begin{table}[width=.9\linewidth,cols=4,pos=h]
\caption{Plasma parameters during passage of solar storm.}\label{tab:Tab1}
\begin{tabular*}{\tblwidth}{@{} LLLLL@{} }
    \toprule
         Stage&Typical&	Shock&	Early CME&	Late CME \\
         \midrule
         Number density($\mathit{\rm /cm^{-3}}$)&	5&	20&	3&	$\geqslant$50 \\
         Bulk speed(km/s)& 450&	600&	650&	500 \\
         $\mathit{T_e}$($\mathit{\rm 10^4}$K)& 9.98&	80.0&	8.00&	3.02 \\
         $\mathit{c_s}$(km/s)& 40.6&	115&	36.4&	22.3 \\
         \bottomrule
\end{tabular*}
\end{table}

Furthermore, a CME emitting from the Sun was considered, both the density and temperature of the solar wind underwent drastic changes, even reaching values 10 times that in usual. This event can be divided into four distinct temporal parts: typical solar wind; a dense, hot shock; early stage of CME; late stage of CME \citep{2012JGRE..117.0K04F}. The four simulation cases and their solar wind parameters have been performed in Table \ref{tab:Tab1}. 

\begin{figure}[ht!]
\centering
\includegraphics[width=0.85\linewidth]{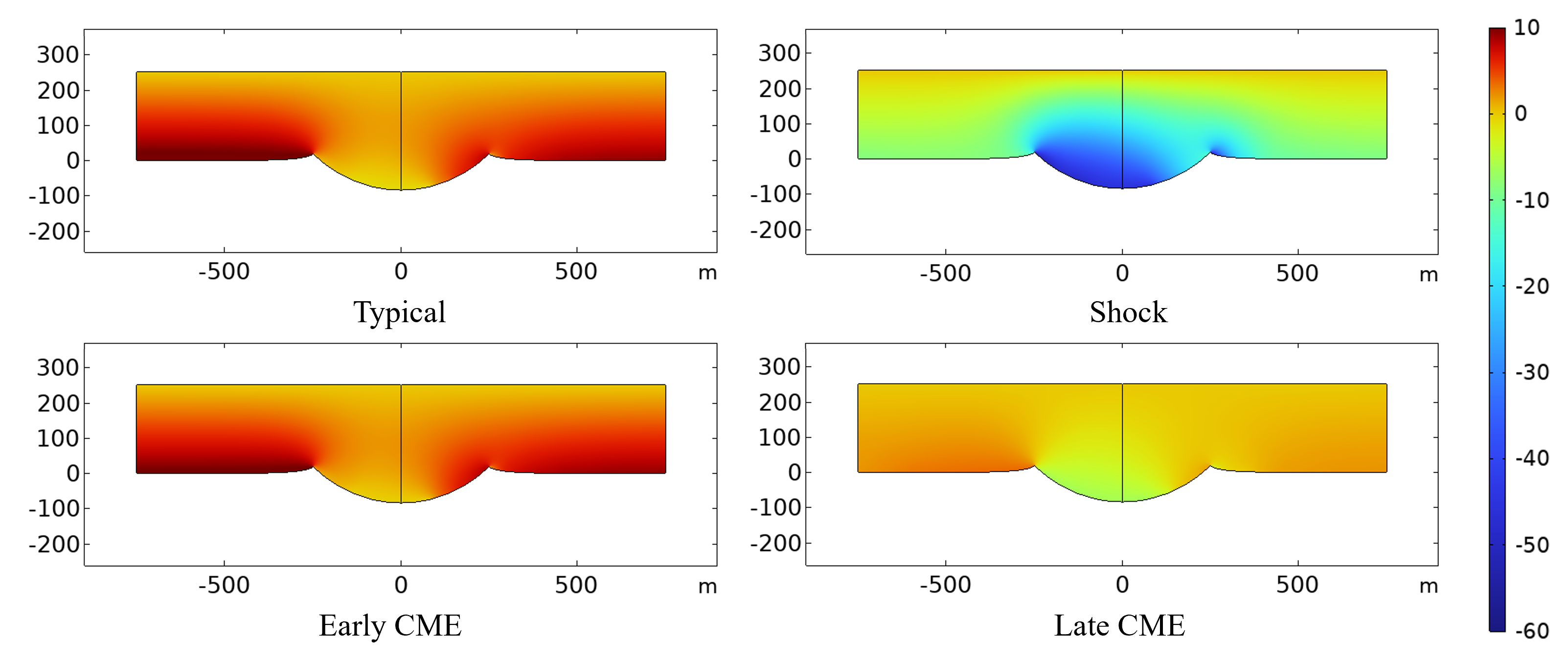}
\caption{Simulated potential during passage of the modeled CME. The crater has a diameter of 500m, and the solar wind is horizontally incident, i.e. $\alpha=0^{\circ}$.
\label{fig:Fig5}}
\end{figure}

\begin{table}[width=.9\linewidth,cols=4,pos=h]
\caption{Surface potentials at the floors of craters with different sizes during various stages of solar storm.}\label{tab:Tab2}
\begin{tabular*}{\tblwidth}{@{} LLLLLLL@{} }
    \toprule
         Stage&$\alpha$&	50m&	200m&	500m&800m \\
         \midrule
         \multirow{3}*{Typical}&	0&-0.97V&	-0.98V&	-0.98V& -0.98V\\
         ~&45&	11.00V&	10.98V&	10.96V&	10.95V  \\
         ~&90&	11.16V&	11.15V&	11.15V&	11.12V  \\
         \midrule
         \multirow{3}*{Shock}& 0&-53.65V&	-57.27V&	-60.11V& -62.78V\\
         ~&45&	-2.44V&	-3.41V&	-5.10V&	-6.79V  \\
         ~&90&	-2.95V&	-4.07V&	-5.72V&	-7.40V  \\
         \midrule
         \multirow{3}*{Early CME}&	0&-0.35V&	-0.36V&	-0.36V& -0.37V\\
         ~&45&	10.26V&	10.21V&	10.13V&	10.05V  \\
         ~&90&	10.33V&	10.28V&	10.19V&	10.12V  \\
         \midrule
         \multirow{3}*{Late CME}&	0&-6.32V&-6.62V&	-7.35V&	-7.78V\\
         ~&45&	4.88V&4.60V&4.04V&	3.97V  \\
         ~&90&	4.89V&4.62V&4.13V&	3.69V  \\
         \bottomrule
\end{tabular*}
\end{table}

As shown in Figure \ref{fig:Fig5}, during the passage of CME, the coupled wake/surface system is highly sensitive to fluctuations in solar wind conditions. When the forward shock occurs, the solar wind number density and temperature experience a sharp increase, and the solar wind starts to accelerate. At this point, a sudden drop can be seen in the surface potential, the photoelectron current is inferior to the current generated by high-energy electrons, the illuminated area becomes negatively charged, reaching -8.41V, while the minimum potential in the shadowed area can drop to -60.11V. The potential difference exceeding 50V results in the exponential growth of the electric field strength, even up to 454.08V/m, greatly enhancing the sensitivity of electrons and ions to the charged properties of the asteroid, making the plasma extremely unstable. Electrons are accelerated to $8.47\times10^8$m/s, accumulating and heating up under the influence of electric field migration. Secondary electron emission, which strongly depends on electron and ion temperatures and is positively correlated with them, is significantly enhanced. In contrast with the other stages of the solar storm, the crater surface generates a substantial number of secondary electrons in the Shock stage. In larger craters, these secondary electrons struggle to escape from the crater, leading to their accumulation at the bottom, further reducing the surface potential. As indicated in Table \ref{tab:Tab2}, the potential at the floor $F$ of the 800-meter crater is reduced by 9.13V compared to that of the 50-meter crater when $\alpha=0^{\circ}$.

Apart from the Shock stage, the potential around $F$ during the Late CME also decreases significantly with increasing crater diameter. The dense, cold late cloud has relatively high electron and ion densities, simultaneously, the extremely low temperature, approximately 0.3 times that under typical solar wind, meaning that the energy of charged particles is insufficient for them to escape from the crater. Electrons, which are more likely to enter the crater than ions, tend to accumulate there along with the photoelectrons and secondary electrons generated at the surface. This phenomenon becomes more pronounced as the depth, i.e. the size of the crater increases.

\subsection{Depth-to-Width Ratio}
\label{subsec:Dep}

\begin{figure}[ht!]
\centering
\includegraphics[width=0.9\linewidth]{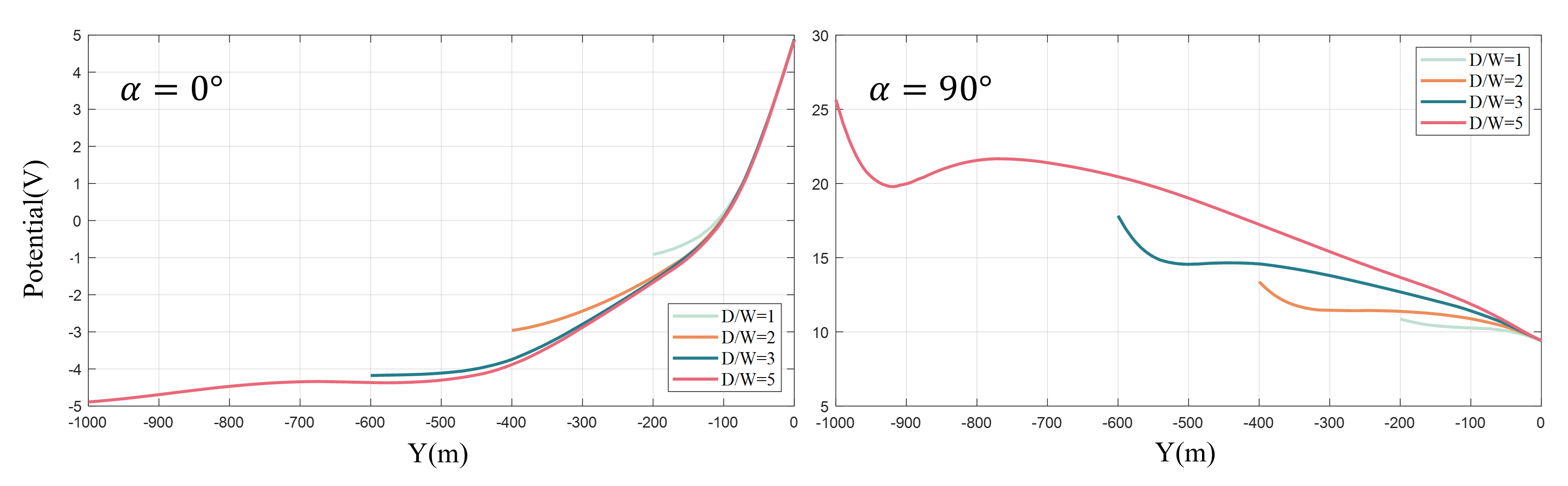}
\caption{The potential distributions along Z-axis penetrating the center of the X-O-Y plane. Craters with different depth-to-width ratios interact with the typical solar wind. The top rim of the crater is at $Y=0$.
\label{fig:Fig6}}
\end{figure}

We simulated the surface charging process for craters with different depth-to-width ratios, each with a diameter of 200m. The results of craters with depth-to-width ratios of 1, 2, 3, and 5 are presented in Figure \ref{fig:Fig6}.

When $\alpha=0^{\circ}$ and the depth-to-width ratio exceeds 1, both electron and ion densities decline drastically inside the crater, with ion density experiencing a more significant reduction. This density feature is attributed to the diverging motions of electrons into the crater and the less deflected trajectories of ions due to the appreciable bulk velocity. Consequently, ions are more likely to skim over the crater due to the horizontal movement of the solar wind or only collide only with the walls near the top rim, making it challenging for them to reach the crater floor. As the depth-to-width ratio increases, the ion density at the crater's bottom decreases exponentially, as explained in Section \ref{subsec:Siz} concerning the canonical theory of quasineutral plasma expansion. Therefore, when the crater is located near the terminator, the surface potential at its bottom  decreases significantly with increasing depth-to-width ratio.

Moreover, this decreasing trend gradually levels off with increasing depth. When the depth exceeds 5 times the width, electrons and ions become extremely thin, falling below 10$\%$ that outside the crater. Further increases in the depth-to-width ratio have a limited effect on the potential at the floor. Under typical solar wind conditions, increasing the depth-to-width ratio from 5 to 10, with an additional depth of 1000m, results in only a 0.33V reduction in potential. During the solar storm, the surface potential within the crater decreases more significantly as the depth-to-width ratio increases, especially in the Shock stage. Due to the enhancement in secondary electron production, the potential can reach -59.58V at a depth-to-width ratio of 0.5, and further decline to -70.98V when the ratio equals 1.

However, when the crater is located at the dayside of the asteroid, where solar wind is incident perpendicularly, i.e. $\alpha=90^{\circ}$, the charging results is completely different. Considering that $v_e>v_b>v_i$ holds with regard to the velocities involved, where $v_e$ and $v_i$ are the velocities of electrons and ions, respectively, $v_b$ is the bulk speed of solar wind, and there exists more divergent nature of motions for electrons than ions, resulting in much greater sticking rate for electrons than ions at the sidewall boundaries of the crater. This suggests that the solar wind ions should be more accessible to the bottoms of deep cavities on the asteroid, leading to a higher concentration of ion deposition at the crater floor. When the depth-to-width ratio becomes sufficiently large (generally exceeding 0.5), the contribution of ions to surface charging at the crater floor surpasses that of incident electrons and secondary electrons, leading to a substantial positive potential. As the depth-to-width ratio increases, the degree of difference between electron and ion motions, specifically their sticking rates at the sidewall and the accessibility to the bottom, is more pronounced \citep{nakazono2023unconventional}. Even during the solar storm, apart from the Shock stage where the electron current remains the dominant charging current, the potential at the crater floor increases with the depth-to-width ratio in the other three stages.

\section{Conclusion} \label{sec:Con}

We studied the surface charging properties of meteorite craters on the asteroid, even the surrounding plasma environment, under various conditions. A simulation method which integrates neural networks and FEM was proposed to implement dynamic three-dimensional simulation, ensuring the reliability and high efficiency of our research. Craters located from the equator to the poles have been modeled and analyzed. It was found that at higher latitudes, because of combined effects of reduced solar radiation and increased shadowing from the crater walls, the potential curve within the crater becomes smooth, eventually stabilizing around -3V when the crater reaches the pole.

Under normal conditions, in addition to the photoelectric effect, the incident electron current and ion current are the primary sources of surface charging. The density difference between them, caused by crater wall shielding, is only related to the ratio of the vertical and horizontal distances from the current position to the crater rim, therefore, the surface potential within the crater is highly sensitive to the depth-to-width ratio, while it shows indifference to changes in the size. When the solar storm is taken into consideration, the energy of electrons and ions changes significantly, and secondary electron emission is enhanced. In large craters, a substantial number of secondary electrons and incident electrons are hard to escape from the crater, leading to their accumulation at the floor, resulting in visible potential differences among craters of various sizes.

For craters with a high depth-to-width ratio, variations in depth can lead to distinctive surface charging results under different solar wind incidence angles $\alpha$, even in the absence of solar storms. For instance, when the solar wind is incident horizontally, i.e. $\alpha=0^{\circ}$, the potential within the crater primarily relies on electrons, decreasing as the depth-to-width ratio increases. Conversely, when $\alpha=90^{\circ}$, ions, primarily driven by the general flow of the solar wind, are less likely to be absorbed by the walls compared to electrons, resulting in a significantly higher ion density at the bottom, which thus exhibits a notably positive potential. This phenomenon intensifies as the depth-to-width ratio increases. It is evident that, apart from the solar wind parameters, the inherent characteristics of the crater also have a significant impact on its surface charging results.

Our research provides simulations of interaction between solar wind plasma and electric field environments around meteorite craters on the asteroid, and conducts detailed analysis on how crater parameters,such as location, size, and depth-to-width ratio, affect the charging results. Studies on the surface charging phenomenon of asteroids featuring intricate terrain are important for future missions, and have general implications for studying the charging properties of other small airless bodies.

\printcredits

\bibliographystyle{cas-model2-names}

\bibliography{cas-sc-template}


\end{document}